\documentclass[aps,pre,twocolumn,amsmath,amssymb,nofootinbib,floatfix,superscriptaddress]{revtex4-1}

\usepackage{amsmath,braket}
\usepackage{amssymb}
\usepackage{amsthm}
\usepackage[dvipdf]{color}
\usepackage{graphicx}
\usepackage{dcolumn}
\usepackage{bm,subfigure}
\usepackage{xcolor,colortbl}
\usepackage[normalem]{ulem}

\newcommand{\xm}[1] {{\color{purple}#1}}

\begin{document}

\title{Geometrically frustrated self-assembly of hyperbolic crystals from icosahedral nanoparticles}

  \author{Nan Cheng}
  \author{Kai Sun}
  \author{Xiaoming Mao}
 \affiliation{
 Department of Physics,
  University of Michigan, Ann Arbor, 
 MI 48109-1040, USA
 }

\begin{abstract} 
Geometric frustration is a fundamental concept in various areas of physics, and its role in self-assembly processes has recently been recognized as a source of intricate self-limited structures.
Here we present an analytic theory of the geometrically frustrated self-assembly of regular icosahedral nanoparticle  based on the non-Euclidean crystal $\{3,5,3\}$ formed by icosahedra in hyperbolic space.  
By considering the minimization of elastic and repulsion energies, we characterize prestressed morphologies in this self-assembly system. Notably, the morphology exhibits a transition that is controlled by the size of the assembled cluster, leading to the spontaneous breaking of rotational symmetry.

\end{abstract}
\maketitle

\section{Introduction}  
Geometrically frustrated assembly, the assembly of building blocks that don't ``fit together'', has become a rich and actively-growing area of both theoretical and experimental research, due to its potential of assembling controllable self-limited assemblies with high complexity, with potential applications in areas from metamaterials, micro-robotics, to biomedical engineering~\cite{sadoc_mosseri_1999,bruss2012non,irvine2010pleats,grason2016perspective,lenz2017geometrical,travesset2017nanoparticle,haddad2019twist,sadoc2020liquid,li2020some,meiri2021cumulative,serafin2021frustrated,hall2023building,hackney2023dispersed}.  Compared to traditional self-assembly systems producing ordered crystalline structures, these geometrically frustrated assembly systems open a new area with unprecedented possibilities.

Various modes of geometric frustration have been studied in the context of geometrically frustrated assemblies, from disk packings on curved surfaces~\cite{irvine2010pleats,modes2007hard}, chiral twisted bundles~\cite{bruss2012non,grason2016perspective}, topological defects of liquid crystals~\cite{ackerman2017diversity}, and polygons/polyhedra which don't tessellate Euclidean space~\cite{lenz2017geometrical,serafin2021frustrated,schonhofer2023rationalizing}, yielding a rich set of morphologies and properties.  In particular, a new theory was proposed in Ref.~\cite{serafin2021frustrated} where non-Euclidean crystals, the ordered ideal  tessellation of polygons and polyhedra at $100\%$ volume fraction in non-Euclidean space, can be used as reference states to explain and potentially predict the self-assembly of geometrically frustrated building blocks in Euclidean space.  This theory has been developed for non-Euclidean crystals in positively curved space: the 3-sphere, $S^3$, which applies to the self-assembly of tetrahedra and potentially also dodecahedra and octahedra.

In this paper we extend this theory of non-Euclidean crystal assembly to negatively curved space---three dimensional (3D) hyperbolic space, $H^3$, and reveal new mathematical structures and resulting stressed morphologies of the assemblies.  In particular, we choose charged regular icosahedral nanoparticles (NPs) as our model system.  This choice is guided by the following facts.  (i) Among of the five regular polyhedra in Euclidean 3D space $E^3$ (tetrahedron, cube, octahedron, dodecahedron, icosahedron), icosahedron only has regular tesselations in $H^3$.  All other polyhedra have tesselations in both $S^3$ and $H^3$ (depending on the number of polyhedra meeting at each edge).  (ii) NPs are versatile self-assembly systems and it has been shown that highly regular icosahedral NPs can be obtained experimentally~\cite{hofmeister1998forty,de2015entropy}.  Following a setup similar to that of Ref.~\cite{serafin2021frustrated} we consider   
the self-assembly of  icosahedral NPs driven by short-ranged attractive interaction and screened electrostatic repulsion. The short-ranged attraction includes both Van der Waals interactions and ligand binding. 
We also consider the the surface energy of the assembled clusters to be low~\cite{yan2019self}. With these assumptions, the NPs tends to form low dimensional structures with morphologies determined by the competition of  stress from the geometric frustration and electrostatic repulsion. 
We show that the hyperbolic crystal $\{3,5,3\}$, which is the lowest stress regular tessellation of icosahedra in $H^3$, provides us with a reference metric $\overline{g}$ of the stress-free state of the icosahedral NPs, characterizing the ground state of the NPs at low temperature (which lies only in $H^3$).  We then consider a slice of this non-Euclidean crystal in Euclidean space $E^3$, where
an energy functional can then be defined using this reference metric, which corresponds to the energy cost by deforming the slice of the non-Euclidean crystal $\{3,5,3\}$ into Euclidean space. We then show how different types of morphologies arise from this geometrically frustrated self-assembly  at different choices of parameters.  In particular, we find that there is a transition from spherical  to cylindrical shells, controlled by the radius of the assembled disk.  Rotational symmetry spontaneously breaks at this transition. 

This theory, developed based on icosahedral NPs, is applicable to general geometrically frustrated self-assembly systems where the stress-free tessellation can be realized in $H^3$.  The general mode of geometric frustration, where ``extra volume'' or ``overlap'' is isotropically generated as the building blocks bind, is partially released by having thin sheets with gaussian curvatures.  We expect this theory to provide guiding principles for future studies of geometrically frustrated assembles characterized by negatively curved non-Euclidean crystal tessellations, which have been predicted to carry exotic excitations~\cite{kollar2019hyperbolic,maciejko2021hyperbolic,yu2020topological,cheng2022band,maciejko2022automorphic,urwyler2022hyperbolic,bzduvsek2022flat}.


\section{Results}
\subsection{Constructing the hyperbolic crystal \{3,5,3\}}
\begin{figure}[h]
    \centering
     \includegraphics[width=0.45\textwidth]{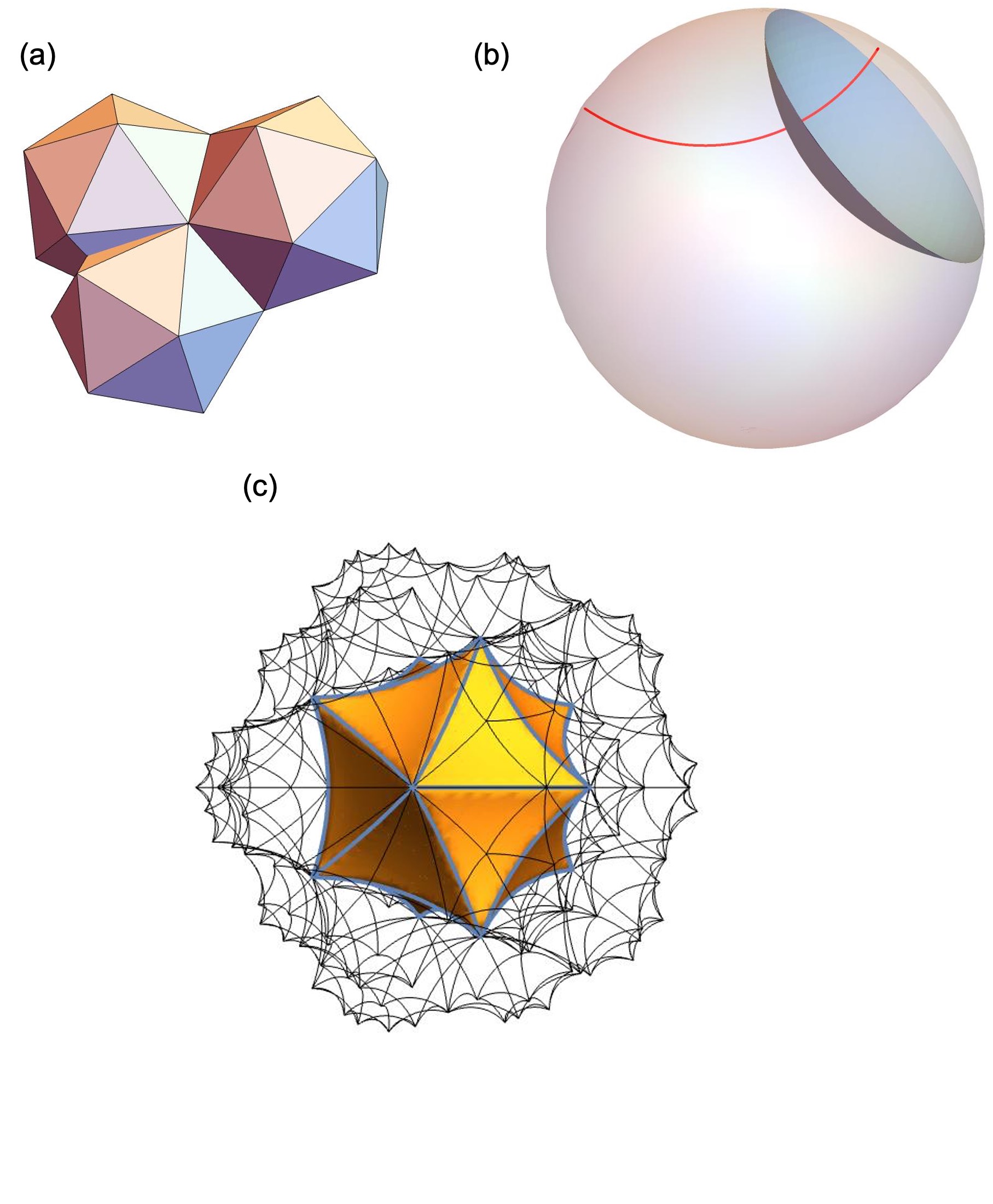}
    \caption{(a) Overlap between 3 icosahedra sharing one edge. (b) An example of a geodesic (red line) and a hyperplane (blue surface) in the Poincar\'{e} ball model. (c) Central hyperbolic icosahedron in the Poincar\'{e} ball model (yellow with thick blue outline) and the first two layers of the hyperbolic crystal \{3,5,3\} (thin black outlines).}
    \label{tiling}
\end{figure}

We start by considering the packing of identical regular icosahedra in space. 
The dihedral angle between two adjacent faces in a regular icosahedron is  $\arccos{\left(-\frac{\sqrt{5}}{3}\right)}\approx 138.19^{\circ}$, three icosahedra sharing a common edge would have a slight overlap of $54.57^{\circ}$ (Fig.~\ref{tiling}a). 
This overlap can be eliminated by introducing negative curvature, i.e., putting the packing in $H^3$, which is an isotropic 3D space with constant negative curvature. One way to intuitively understand this is by realizing that
negative curvature can enlarge the space because a sphere of radius $r$ has volume proportional to $e^r$ in $H^3$ instead of $r^3$ in $E^3$. Therefore, negatively curved space can accommodate more volume, making it possible to have 3 icosahedra sharing a common edge without overlap.

Among all crystals that can be tessellated by icosahedra, the hyperbolic crystal \{3,5,3\} is the ``least frustrated'', meaning that the $H^3$ it corresponds to has the smallest negative curvature (Fig.~\ref{tiling}c). The notion of \{3,5,3\} follows the Schl\"{a}fli symbol $\{p,q,r\}$ for regular tessellations, where one starts with regular polyhedra with $q$ $p$-sided regular polygons meet at each vertex, and fits these polyhedra together such that $r$ of them meet around each edge, forming the tessellation.

As detailed in App.~\ref{APP:hyperboloid} and \ref{APP:353}, we construct the \{3,5,3\} non-Euclidean crystal by mirror reflections starting from the fundamental domain of the icosahedron, following the Coxeter group~\cite{coxeter1999beauty}.  The resulting crystal is infinitely large.  To perform calculations and visualize the results, we adopt the Poincar\'{e} ball model for $H^3$ (detailed in App.~\ref{APP:Poincare}). 
The Poincar\'{e} ball is an open unit ball in $\mathbb{R}^3$ together with the metric
\begin{equation}
ds^2=4\frac{dx^2+dy^2+dz^2}{(1-x^2-y^2-z^2)^2} .
\end{equation}
From this metric it is clear that distances between points close to the boundary of the Poincar\'{e} ball diverges,  bringing points at $\infty$ in $H^3$ to this unit ball. 
In the Poinca\'{e} ball model, a geodesic (i.e., a length-minimizing curve between two points) is an arc perpendicular to the boundary of the unit ball, and a  hyperplanes (i.e., surfaces with constant negative curvature that can be decomposed into disjoint union of geodesics) are spheres perpendicular to the boundary of the unit ball (Fig.1b). 

The hyperbolic tiling \{3,5,3\} can be constructed via hyperbolic reflections, similar to how Euclidean crystals can be generated by mirror reflections in Euclidean space.
First, construct a hyperbolic icosahedron (i.e., a polytope in $H^3$ with 20 faces having the same symmetry as icosahedron in Euclidean space) centered at the origin.  The curvature of this $H^3$ is chosen such that the  dihedral angle between two adjacent faces is $\frac{2\pi}{3}$ (Fig. 1c).  Then, reflect the central icosahedron using its 20 faces, we get the second layer of icosahedron. Repeating this process we get the entire tiling (Fig. 1c, and as detailed in App.~\ref{APP:353}).

\subsection{Model energy of frustrated NP self-assembly}
We next consider the energy of the assemblies as one ``flattens'' the non-Euclidean crystal \{3,5,3\} to our Euclidean space.  
We consider a model energy of four parts~\cite{serafin2021frustrated},
\begin{equation}
E=E_{elastic}+E_{repulsion}+E_{binding}+E_{boundary}
\end{equation}
Here $E_{binding}$ describes the binding energy between the NPs, $E_{boundary}$ describes the exposed surfaces of the NPs in contact with the solvent.  We consider the regime where $E_{binding}$ is large (e.g., $\sim 50k_B T$ between a pair of bound NPs), and thus tight face-to-face binding following the pattern of the \{3,5,3\} non-Euclidean crystal is favored.
Therefore, we can assume the assembly is obtained by choosing a suitable slice $\overline{M}$ of the 
hyperbolic crystal $\{3,5,3\}$, and flattening it into the Euclidean space.
Once $\overline{M}$ is chosen, $E_{binding}$ and $E_{boundary}$ are fixed, as $E_{boundary}$ is determined by the amount of surfaces exposed to the solution and $E_{binding}$ depends  on the number of attached surfaces.

The morphology of $\overline{M}$ in our Euclidean space is determined by the combination of $E_{elastic}$ and $E_{repulsion}$. We take the continuum limit to enable an analytic treatment of the problem.  The elastic energy comes from the fact that the \{3,5,3\} non-Euclidean crystal necessarily needs to be distorted to fit into $E^3$.  
Let $g_{\mu}^{\tau}$ be the actual Euclidean metric of the assembly and $\overline{g}_{\mu}^{\tau}$ be the curved metric of $H^3$ representing the ideal, stress-free distance between the NPs. In the continuum theory, close to a local minimum the elastic energy can be generally written as that of an isotropic homogeneous elastic medium of Lam\'{e} coefficients $\lambda,\mu$,
\begin{equation}
E_{elastic}=\int_{\overline{M}}\left\lbrack\lambda(\epsilon_{\nu}^{\nu})^2+\frac{1}{2}\mu\epsilon_{\nu}^{\tau}\epsilon_{\tau}^{\nu}\right\rbrack
 \sqrt{\det{\overline{g}}}\, d^3x ,
\end{equation}
where $\sqrt{\det{\overline{g}}}\, d^3x$ is the volume element in $H^3$, $\epsilon$ is the strain tensor defined as
\begin{equation}
\epsilon_{\mu}^{\tau}=\frac{1}{2}(g_{\mu}^{\tau}-\overline{g}_{\mu}^{\tau}),
\end{equation}
where $\overline{g}_{\mu}^{\tau}$ and  $g_{\mu}^{\tau}$ can not be equal as the former is a curved metric and the later is a flat metric. Hence, stress necessarily develops in the self-assembly process.

The energy $E_{repulsion}$ characterizes electrostatic repulsion between these charged NPs, using screened Coulomb potentials,
\begin{equation}\label{coulombscreened}
V(r_1,r_2)=\frac{1}{4\pi\epsilon_0}\frac{q_1q_2}{|r_1-r_2|}e^{-\kappa|r_1-r_2|} .
\end{equation}

Since $E_{binding}$ and $E_{boundary}$ are determined by the part $\overline{M}$ we choose in the non-Euclidean crystal $\{3,5,3\}$, $E_{elastic}$ and $E_{repulsion}$ are determined by how this part $\overline{M}$ is embedded in the Euclidean space. We minimize the total energy in two steps. First, we determine the appropriate part $\overline{M}$. Second, we numerically minimize the repulsion and elastic energy in the space of all possible morphologies of $\overline{M}$.


\subsection{Choosing a slice of $\{3,5,3\}$ and the thin shell expansion}
In general, choosing the right slice $\overline{M}$ from the non-Euclidean crystal $\{3,5,3\}$ to predict the self-assembly in Euclidean space is a highly nontrivial question, as the real self-assembly process is out of equilibrium with complex kinetic pathways that determine the assemblies.  

As a simple model, we consider a smooth 2D slice $\overline{M}$ which has zero gaussian curvature in $H^3$.  The general theory we present here, in terms of constructing the non-Euclidean crystal and minimizing the model energy, is applicable to other choices of the slice. 
The choice of $\overline{M}$ we take here is guided by the  following considerations. 
First, given the electrostatic repulsion and the low surface energy, 
the NPs prefer to form 2D structures. 
Second, for the scalability of the assembly process, the 2D structure should have constant Gaussian curvature (otherwise pieces of this sheet can't merge). This gives us three cases: 2D surface with constant negative, positive or zero Gaussian curvature, all cut from the non-Euclidean crystal $\{3,5,3\}$. 

\begin{figure}[h]
    \centering
    \includegraphics[width=0.45\textwidth]{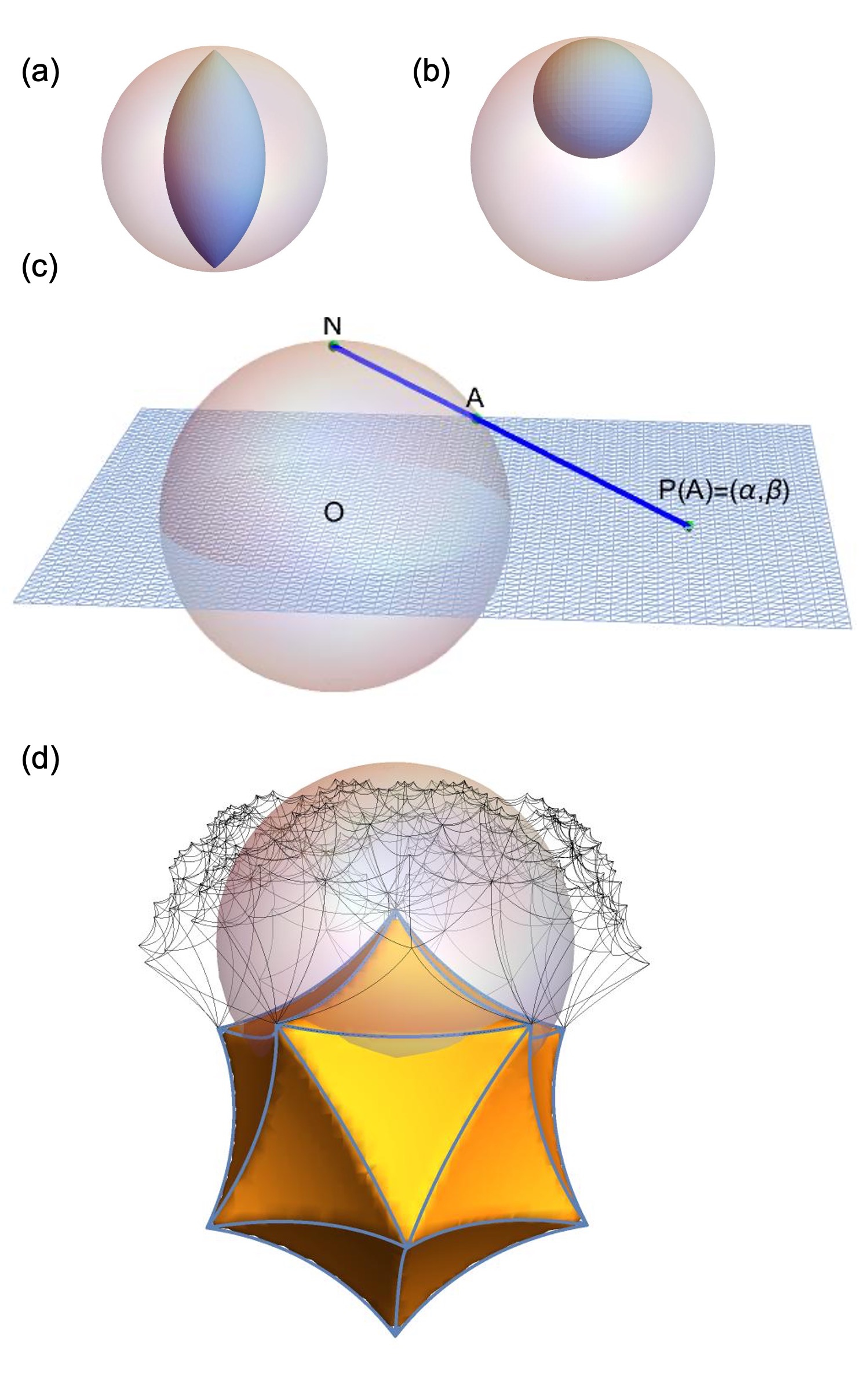}
    \caption{(a) A surface equidistant to a geodesic ($z$ axis) in the Poincaré ball model. (b) A horosphere in the Poincaré ball model. (c) Stereographic projection of the horosphere around the north pole, defining the coordinate $\alpha, \beta$. (d) The subset of icosahedra in \{3,5,3\} that intersect with the horosphere at $\delta=0$ (the white sphere).}
    \label{flatsurface}
\end{figure}

For surface with constant negative Gaussian curvature, the area of a disk of radius $r$  is proportional to $e^r$. Embedding such a surface into Euclidean space would cost too much elastic energy since the area a disk of radius $r$ in Euclidean space only grows as $r^2$. Hence, such a surface is not energetically preferred.
There are two types of surfaces with zero Gaussian curvature: surfaces equidistant to a geodesic and horospheres (Fig.~\ref{flatsurface}a and b). Horospheres are all spheres tangent to the boundary of the Poincaré ball when viewed in the Poincaré ball model. These two surfaces give us the same continuous theory when the number of icosahedra is large (see App.~\ref{APP:ab}). Hence we only consider the horosphere.
For surfaces with constant positive Gaussian curvature, after a hyperbolic translation, they can be thought of as spheres centered at origin. However, the curvature of such a sphere decays exponentially as a function of its radius. 
The theory thus essentially reduces to the zero Gaussian curvature case (see App.~\ref{APP:ab}) when we consider a large surface where the continuum theory is appropriate.

\begin{figure*}[t]
    \centering
    \includegraphics[width=\textwidth]{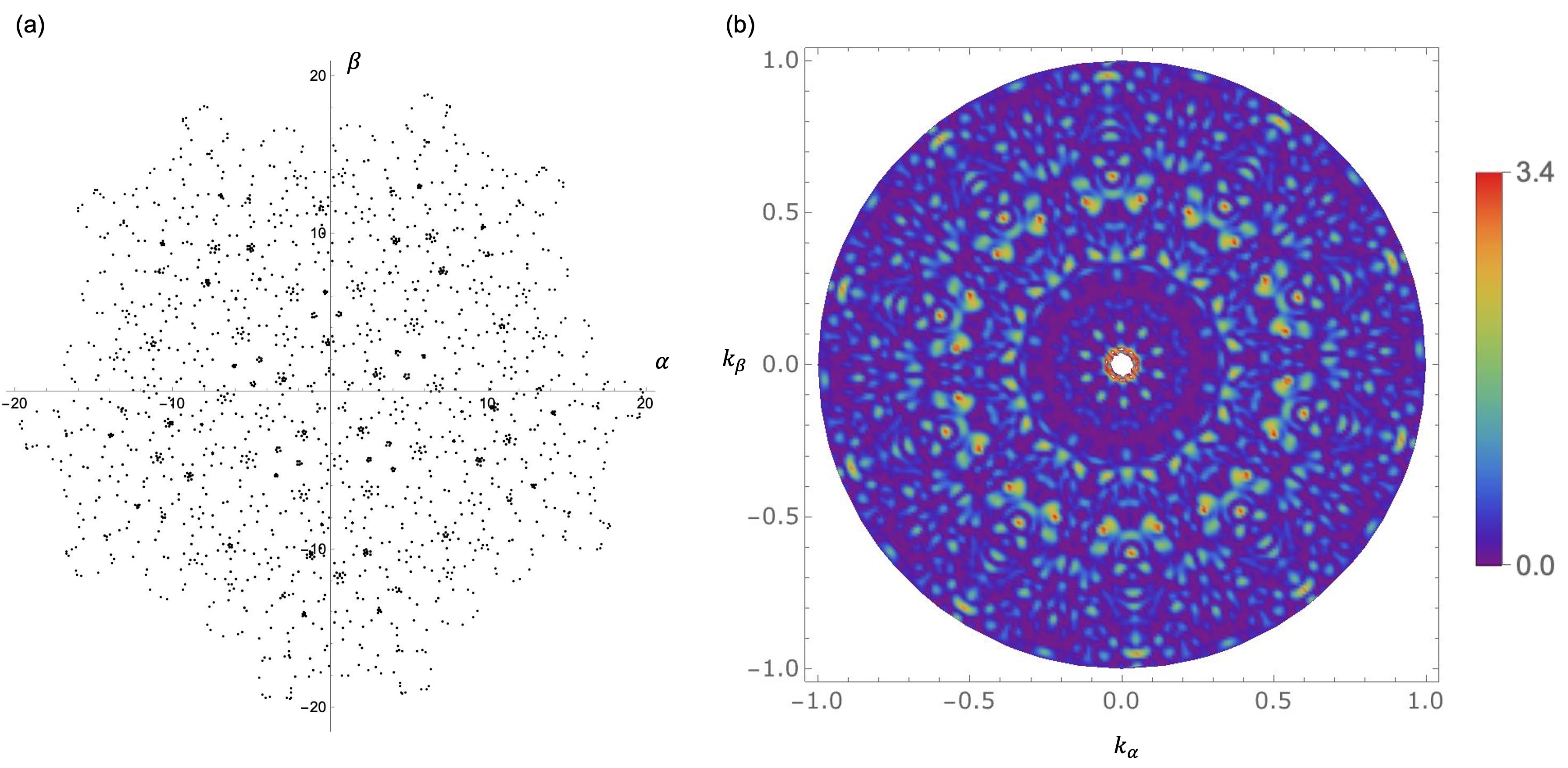}
    \caption{(a) The intersection between the horospheres and the edges in the $\{3,5,3\}$ tiling up to the fourth layer. (b) The diffraction pattern of the interaction points.}
    \label{quasicrystal}
\end{figure*}

\begin{figure*}[h]
    \centering
     \includegraphics[width=1.0\textwidth]{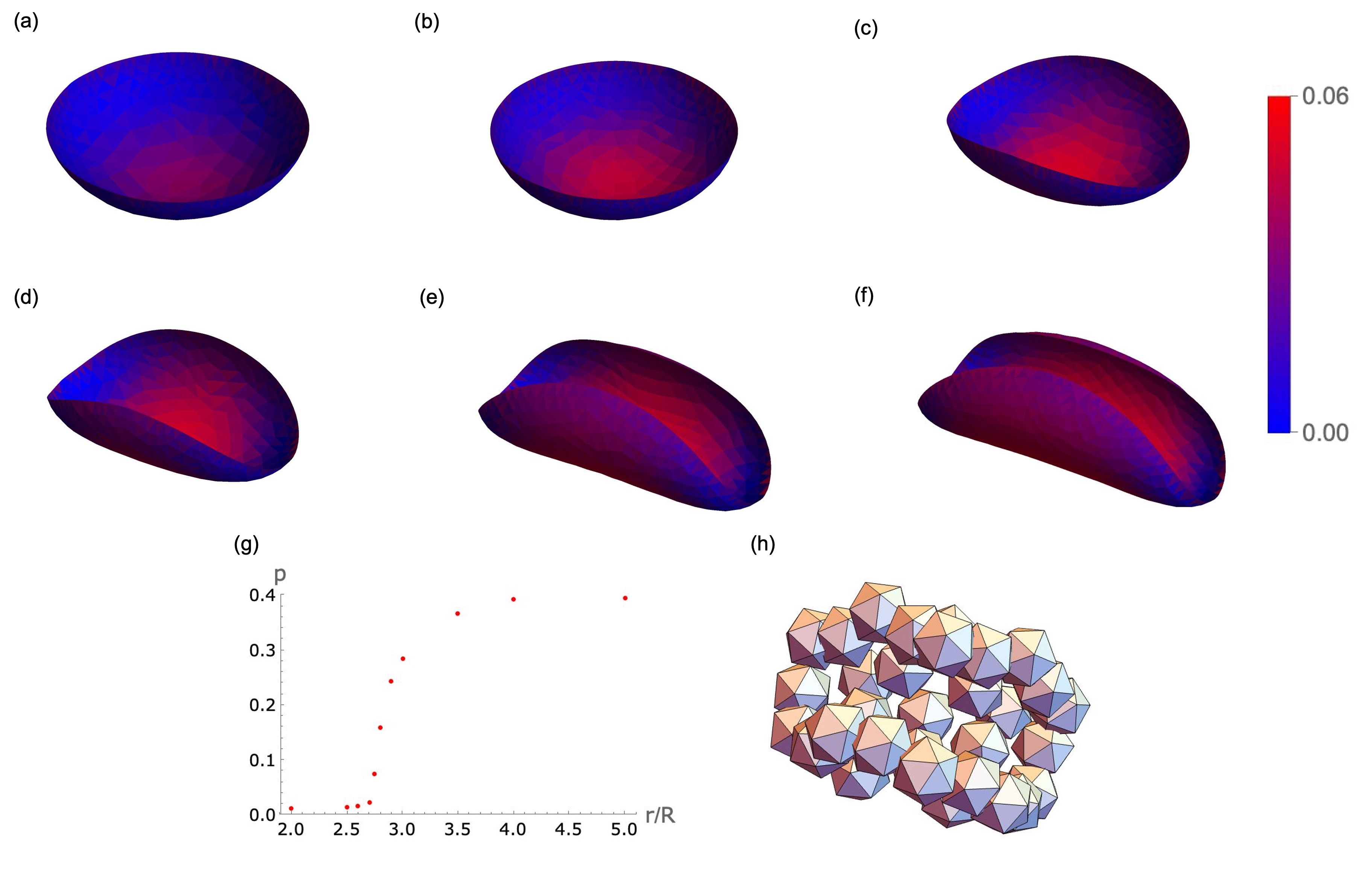}
    \caption{(a)-(f) Mid-surface morphology and mid-surface elastic energy density  of disks of radius 2 (a), 2.5 (b), 2.8 (c), 3 (d), 4 (e), 5 (f) respectively, the red (blue) region denotes high (low) elastic energy density. (g) The symmetry breaking parameter, zero means no rotational symmetry breaking. (h) Actual position of icosahedra in the minimal energy configuration \xm{at radius 5}.}
    \label{morphology}
\end{figure*}


We next formulate a mathematical description of a horosphere for this problem.  To do this, we first use the following $\{\alpha, \beta,\delta\}$ coordinate system to parameterize the Poincaré ball,
\begin{equation}
\begin{aligned}
&x=\frac{2\alpha(\cosh\delta+\sinh\delta)}{2+(2+\alpha^2+\beta^2)\cosh\delta+(\alpha^2+\beta^2)\sinh\delta},\\
&y=\frac{2\beta(\cosh\delta+\sinh\delta)}{2+(2+\alpha^2+\beta^2)\cosh\delta+(\alpha^2+\beta^2)\sinh\delta},\\
&z=\frac{(\alpha^2+\beta^2)\cosh\delta+(\alpha^2+\beta^2-2)\sinh\delta}{2+(2+\alpha^2+\beta^2)\cosh\delta+(\alpha^2+\beta^2)\sinh\delta}.\\
\end{aligned}
\end{equation}
Here $\{\alpha, \beta,\delta\}$ is the coordinate system (i.e., they label the icosahedra), and $\{x,y,z\}$ is the embedding of these points in $H^3$, represented using the Poincaré ball model.
Under this parameterization, any constant $\delta$ surface is a horosphere tangent to the boundary of the Poincaré ball at $(0,0,1)$ and the $\delta$ direction can be viewed as thickness direction of the horospheres. 
The $\{\alpha, \beta\}$ coordinates are chosen such that they coincide with the stereographic projection of the horosphere at each given $\delta$ from the north pole to a plane normal to $z$. 
The metric tensor $\overline{g}$ under this coordinate system is
\begin{equation}
\overline{g}=
\begin{pmatrix}
1 & 0 & 0\\
0 & e^{2\delta} & 0\\
0 & 0 & e^{2\delta}
\end{pmatrix} .
\end{equation}
Because horospheres at different $\delta$ are equivalent (the Poincaré ball represents infinite $H^3$), without losing generality, 
we choose $\delta=0$ surface which passes through the center of the Poincaré ball as the center surface of $\overline{M}$ and expand $\overline{g}(\delta)$ around $\delta=0$,
\begin{equation}
\overline{g}_{ij}(\delta)=\overline{a}_{ij}-2\delta\overline{b}_{ij}+\delta^2\overline{c}_{ij}+O(\delta^3) ,
\end{equation}
where the reference first and second fundamental form
\begin{equation}\label{referencefundamentalforms}
\overline{a}=
\begin{pmatrix}
1 & 0\\
0 & 1
\end{pmatrix}
,\quad \overline{b}=
\begin{pmatrix}
-1 & 0\\
0 & -1
\end{pmatrix} .
\end{equation}
It is worth noting that different choice of $\delta$ gives us different reference first and second fundamental form but they only differ by a scaling factor and don't affect the embedding.

From the form of $\overline{a},\overline{b}$, it is clear that the reference in-plane metric of this horosphere is flat (as it is of zero Gaussian curvature), but the sheet expand/shrink at different heights across the normal direction, which is incompatible with the flat in-plane metric.  This is the mode of geometric frustration in this problem.

This surface then intersect the icosahedra in the $\{3,5,3\}$ crystal, which will be the material in the assembly. 
Fig.~\ref{flatsurface}d shows the icosahedra intersecting with the $\delta=0$ horosphere in the Poincaré sphere. It is worth noting that although in the figure it appears to be a finite sphere intersecting distorted icosahedra of different sizes, the actual picture of this assembly in $H^3$ is infinitely large, and all icosahedra are regular and identical, although they intersect the horosphere at different heights. 

The intersection between the horosphere and the icosahedron edges are shown in Fig.~\ref{quasicrystal}a. These points have 5-fold rotational symmetries as the \{3,5,3\} tiling does. One interesting observation is that this pattern 
may be defined as a type of quasicrystal, which is new to our knowledge.  This can be seen by doing a Fourier transform of the edge intersection pattern in Fig.~\ref{quasicrystal}a, which results in Fig.~\ref{quasicrystal}b.  The sharp diffraction peaks and the 5-fold rotational symmetry, as well as the aperiodicity of the real space pattern, satisfies the criterion of ``quasicrystals''.  Distinct from known quasicrystals, some of which can be obtained by cutting a high dimensional (Euclidean) crystal using a low dimensional surface, this is obtained by cutting a non-Euclidean crystal using a 2D surface with no Gaussian curvature.

When the thickness is much smaller than the in-plane dimension of the surface and the radius of curvature (here we choose suitable unit of length such that the radius of curvature is one). We can apply thin shell approximation and the elastic energy density is
\begin{equation}
\begin{aligned}
\epsilon_{s}=&\frac{hY}{8(1-\nu^2)}\{\nu [(a-\overline a){^i_i}]^2+(1-\nu)(a-\overline a){^i_j}(a-\overline a){^j_i}\} ,\\ 
\epsilon_{b}=&\frac{h^3Y}{12(1-\nu^2)}\{\nu [(b-\overline b){^i_i}]^2+(1-\nu)(b-\overline b){^i_j}(b-\overline b){^j_i}\} ,
\end{aligned}
\end{equation}
where $Y$ is the Young's modulus, $h$ is the thickness, $\nu$ is the Poisson's ratio, $\epsilon_{s}$ is the stretching energy density and $\epsilon_{b}$ is the bending energy density. The total elastic energy under this thin shell expansion is
\begin{equation}\label{elasticenergy}
E_{elastic}=\int_{\overline{M}} d\alpha d\beta \sqrt{\overline{a}}(\epsilon_{s}+\epsilon_{b}) .
\end{equation}
Under the same thin shell approximation, the electrostatic repulsion energy can be written as,
\begin{equation}\label{repulsionenergy}
E_{repulsion}=\int_{p\in\overline{M}}\int_{q\in\overline{M}}\frac{1}{8\pi\epsilon_0}\frac{\rho^2 dS_1dS_2}{|\vec{r}(p)-\vec{r}(q)|}e^{-\kappa|r(p)-r(q)|}.
\end{equation}
where $\rho$ is the charge density, $\vec{r}(p)$ is the position of $p\in\overline{M}$ in the Euclidean space $\mathbb{E}^3$, 
$dS_1$ and $dS_2$ are area elements of $\overline{M}$ around $p$ and $q$ respectively.

\subsection{Minimizing the total energy and predicting the morphology}
We next choose suitable parameters and computationally minimize $E_{elastic}+E_{repulsion}$ for these thin shells.  
In the \{3,5,3\} tiling, the radius of the hyperbolic icosahedron is 0.88 (in unit of radius of curvature), slightly smaller than the radius of curvature. We choose $\overline{M}$ to be a disk of radius ranging between 2 to 5 on the horosphere, reasonably greater than the radius of the hyperbolic icosahedron, so that the thin shell approximation is justified.  

We numerically minimize $E_{elastic}+E_{repulsion}$ (see App.~\ref{APP:miniminazation}) of this thin shell.  We have chosen the interaction parameters to be $h=R$ (thickness of the shell), $\rho=\frac{8\sqrt{Y\epsilon_0}}{9\sqrt{\pi}}$ (charge density), $\kappa=\frac{1}{R}$ (inverse screening length), where $R$ is the radius of curvature of the hyperbolic space, and we have expressed the charge density in terms of the relative strength between the repulsion and the elasticity.
These choices are guided by the requirement of making the elastic energy to be of the same order of magnitude of the repulsion energy, so that the balance of the two leads to nontrivial morphologies.  The methods described here are not limited to this regime.


A series of morphologies we obtain at different disk sizes are shown in Fig.~\ref{morphology}a-f.  
It is worth noting that although the reference fundamental forms $\overline{a}$ and $\overline{b}$ are rotationally symmetric, the final morphology can spontaneously  break this symmetry. 
In particular, disks at small radius keep this symmetry, whereas larger disks become more cylinderical.  This results from a crossover from bending dominated regime to stretching dominated regime as the radius of the disk becomes larger. 
Fig.~\ref{morphology}g shows the deviation from rotational symmetric morphology as a function of radius. To quantify this effect, we introduce the following order parameter $P$ to measure the rotational symmetry breaking of the surface. For a given point $\vec{r}$ on a surface $U$, let $\vec{N}_{\vec{r}}$ be the normalized normal vector of the surface at $\vec{r}$. We first average $\vec{N}_{\vec{r}}$ over $U$ to find the rotation vector $\vec{k}$.  Then we choose the rotation axis to be a line $l$ aligning with the rotation vector going through the mass center $\vec{r}_c$ of the surface.  Next, let $R(\theta,l)$ be the the rotation operator about $l$ by angle $\theta$, and we define the order parameter as
\begin{equation}
    P \equiv \int_U \Big{\vert}\frac{d R(\theta,l)\vec{r}}{d\theta}\cdot \vec{N}_{\vec{r}}\Big{\vert}dS = \int_U \Big{\vert}[\vec{k}\times(\vec{r}-\vec{r}_c)]\cdot \vec{N}_{\vec{r}}\Big{\vert}dS
\end{equation}
where $dS$ is the area element of $U$.  When $U$ has rotational symmetry, $\vec{k}\times(\vec{r}-\vec{r}_c)]$ is normal to $\vec{N}_{\vec{r}}$ everywhere, leading to $P=0$.  Nonzero $P$ captures how strongly the morphology breaks this symmetry. 

Fig.~\ref{morphology}h shows the morphology of the assembly when the icosahedra are explicitly shown. They are placed following their positions in the hyperbolic crystal, as discussed in App.~\ref{APP:miniminazation}. Although these icosahedra are not in perfect face-to-face binding, our calculation shows that the assembly minimizes strain and repulsion by adopting this morphology. 


\section{Conclusion}
In this paper, we show that the self assembly of small cluster of charged icosahedral NPs under face-to-face attractive interaction and low surface energy can be understood using the hyperbolic crystal $\{3,5,3\}$. By choosing zero-Gaussian curvature slices of the hyperbolic crystal $\{3,5,3\}$ and minimizing their energies in the Euclidean space, we find morphologies of these assembly, characterized by a transition where rotational symmetry breaks.  
We also find a possible new type of quasicrystal by cutting the hyperbolic crystal \{3,5,3\} using the horosphere. 

This theory provides a new framework to employ non-Euclidean crystals beyond those residing in the 3-sphere, to predict the self-assembly of hyperbolic structures.  These structures may be of interest both as new types of assemblies at the nanoscale, but also for emergent phenomena in hyperbolic space.

\noindent{\it Acknowledgements}--- The authors acknowledge helpful discussions with Francesco Serafin, Nicholas Kotov, Sharon Glotzer, and Philipp Schönhöfer.  
This work was supported in part by the Office of Naval Research (MURI N00014-20-1-2479).  X.M. acknowledges the hospitality of the Kavli Center for Theoretical Physics, supported by the National Science Foundation (NSF PHY-1748958), where this manuscript was completed.




\appendix

\section{The hyperboloid model of hyperbolic space}\label{APP:hyperboloid}
In this section we review the hyperboloid model of hyperbolic space and hyperbolic reflections in the hyperboloid model~\cite{iversen1992hyperbolic}. 

Let $(,)$ be a symmetric bilinear form on $\mathbb{R}^4$ defined by
\begin{equation}
(u,v)\equiv u^Tg_0v
\end{equation}
where
\begin{equation}
g_0=\begin{pmatrix}
-1 & 0 & 0 & 0\\
0 & 1 & 0 & 0\\
0 & 0 & 1 & 0\\
0 & 0 & 0 & 1\\
\end{pmatrix}
\end{equation}
is the Minkowski metric tensor.  

The hyperboloid model of hyperbolic space consists of a submanifold of $\mathbb{R}^4$
\begin{equation}
H^3=\{u=\{t,x,y,z\}\in\mathbb{R}^4, (u,u)=-1, t>0\}
\end{equation}
and a metric $\overline{g}$ on $H^3$ induced by $g_0$, where $\overline{g}(v,v)=v^Tg_0v$ for any vector $v\in\mathbb{R}^4$ tangent to $H^3$. The metric $\overline{g}$ gives a distance function $d$ on $H^3$ satisfying
\begin{equation}
\cosh{d(p,q)}=-(p,q) .
\end{equation}
An isometry of $H^3$ is a smooth map $A: H^3\rightarrow H^3$ preserving distance:
\begin{equation}
d(A(p),A(q))=d(p,q), \forall p,q\in H^3 ,
\end{equation}
which is equivalent to
\begin{equation}\label{isometry}
(A(p),A(q))=(p,q), \forall p,q\in H^3 .
\end{equation}
Form \eqref{isometry}, we see that any linear map $L: \mathbb{R}^4\rightarrow \mathbb{R}^4$ whose matrix $L$ under the standard basis of $\mathbb{R}^4$ satisfying $L^Tg_0L=g_0$ induces a smooth map 
\begin{equation}
\begin{aligned}
A_L: &H^3\rightarrow H^3\\
&\;x\mapsto Lx
\end{aligned}
\end{equation}
and the induced map $A_L$ is an isometry. It turns out that the correspondence $L\mapsto A_L$ is bijective. In the following, we identify a matrix $L$ satisfying $L^Tg_0L=g_0$ with an isometry $A_L$ of $H^3$.

For any vector $v\in\mathbb{R}^4$ satisfying $(v,v)=1$, consider the linear map
\begin{equation}
R_v(x)\equiv x-2(x,v)v .
\end{equation}
$R_v$ satisfies
\begin{equation}
(R_v(p),R_v(q))=(p,q), \forall p,q\in H^3
\end{equation}
Therefore, $R_v$ is an isometry of $H^3$. Since we have 
\begin{equation}
\begin{aligned}
&(R_v)^2=1\\
&R_v(v)=-v\\
&R_v(x)=x, \forall (x,v)=0
\end{aligned}
\end{equation}
$R_v$ can be viewed as a reflection about the hyperplane $P=\{x\in\mathbb{R}^4, (x,v)=0\}$ and such $R_v$ is called a hyperbolic reflection~\cite{iversen1992hyperbolic}. It is worth noting that a hyperbolic reflection can be identified with its normal vector $v$ or its reflecting plane $P$.

\begin{figure}[h]
    \centering
    \includegraphics[width=0.4 \textwidth]{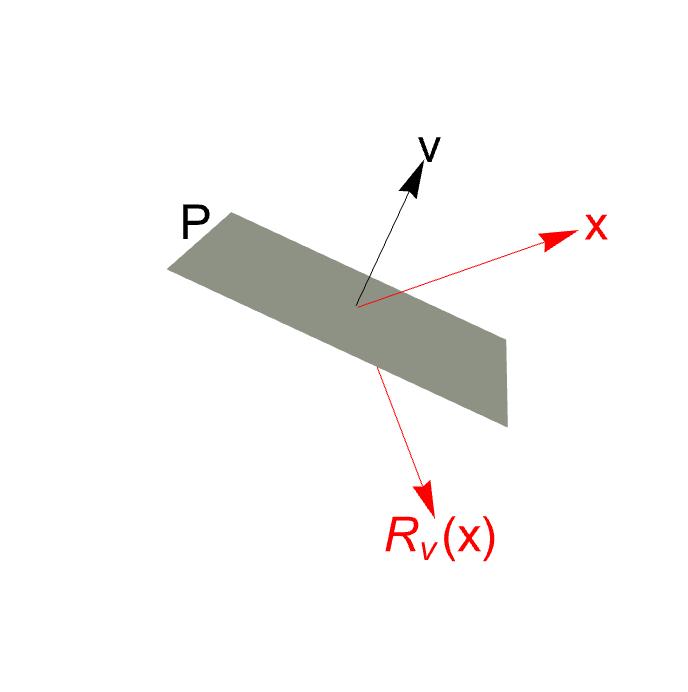}
    \caption{Schematic picture of a hyperbolic reflection.}
    \label{hyperbolicreflection}
\end{figure}

\section{Constructing the \{3,5,3\} crystal in the hyperboloid model}\label{APP:353}
In this section, we use hyperbolic reflections to construct hyperbolic crystal $\{3,5,3\}$. There are three steps in the construction. First, construct a pyramid $K$ that is $\frac{1}{120}$ of one hyperbolic icosahedron in $\{3,5,3\}$. Second, reflect $K$ by three of its four surfaces to form the central hyperbolic icosahedron $M$. Third, reflect $M$ by its surfaces and we get the entire crystal.\\

\begin{figure}[h]
    \centering
    \includegraphics[width=0.4 \textwidth]{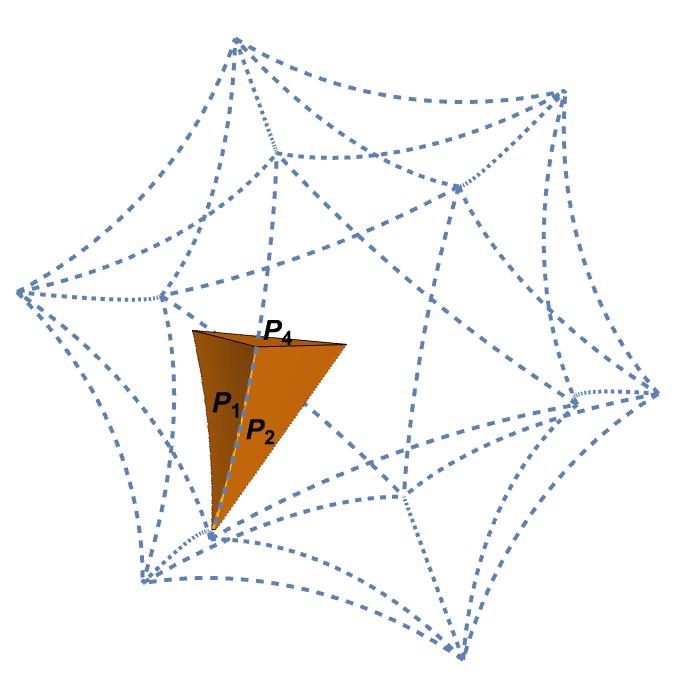}
    \caption{The region K used in generating the hyperbolic icosahradon and the $\{3,5,3\}$ crystal. }
    \label{fundamentalregion}
\end{figure}

We first construct the pyramid $K$. Let $P_1, P_2, P_3, P_4$ be the four surfaces of $K$, $v_1, v_2, v_3, v_4$ be the corresponding normal vectors under $g_0$ and $R_{v_1}, R_{v_2}, R_{v_3}, R_{v_4}$ be the corresponding hyperbolic reflections. In the $\{3,5,3\}$ tiling, $R_{v_1}, R_{v_2}, R_{v_3}, R_{v_4}$ satisfies
\begin{equation}
(R_{v_i}R_{v_j})^{m_{ij}}=1
\end{equation}
where
\begin{equation}
m=\begin{pmatrix}
2 & 3 & 2 & 2\\
3 & 2 & 5 & 2\\
2 & 5 & 2 & 3\\
2 & 2 & 3 & 2\\
\end{pmatrix}
\end{equation}
The corresponding vectors satisfies
\begin{equation}
(v_i,v_j)=-\cos{\frac{\pi}{m_{ij}}}
\end{equation}
$v_i$'s are unique up to global rotations and we can take them to be
\begin{equation}
\begin{aligned}
&v_1=(\sqrt{\frac{16\cos^2(\pi/5)-9}{12-16\cos^2(\pi/5)}},-\sqrt{\frac{3}{12-16\cos^2(\pi/5)}},0,0)^T\\
&v_2=(0,\sqrt{1-\frac{4\cos^2(\pi/5)}{3}},-\frac{2\sqrt{3}\cos(\pi/5)}{3},0)^T\\
&v_3=(0,0,\frac{\sqrt{3}}{2},- \frac{1}{2})^T\\
&v_4=(0,0,0,1)^T
\end{aligned}
\end{equation}
Then we move $K$ by the group $S$ (a group of order 120) generated by $R_{v_2}, R_{v_3}, R_{v_4}$ to form the central hyperbolic icosahedron $M$. The 20 surfaces of $M$ corresponds to reflections $\{sR_{v_1}s^{-1}, s\in S\}$. Finally, we use the 20 surfaces of $M$ to reflect $M$ and get the entire crystal~\cite{coxeter1999beauty}.

\section{The relation between the hyperboloid model and the Poincaré ball model}\label{APP:Poincare}
The hyperboloid model reviewed above is convenient for calculations of hyperbolic reflections, but the Poincaré ball model, which we use in the main text, is more convenient to visualize the hyperbolic space~\cite{iversen1992hyperbolic}.  Here we review their relations.

The Poincaré ball model consists of an open unit ball
\begin{equation}
B=\{\{x,y,z\}\in\mathbb{R}^3,x^2+y^2+z^2<1\}
\end{equation}
in $\mathbb{R}^3$ and a metric
\begin{equation}
ds^2=4\frac{dx^2+dy^2+dz^2}{(1-x^2-y^2-z^2)^2}
\end{equation}
If we identify the Poincaré ball with $\{\{0,x,y,z\}\in\mathbb{R}^4,x^2+y^2+z^2<1\}\subset\mathbb{R}^4$, it is then related to the hyperboloid model projectively. If we have a point $\{t,x,y,z\}$ in the hyperboloid model, we may project it onto the $t=0$ hyperplane by intersecting it with a line drawn through $\{-1,0,0,0\}$. The result is the corresponding point of the Poincaré disk model. To put it explicitly, the projection $\pi: H^3\rightarrow B$ sends $\{t,x,y,z\}\in H^3$ to $\{\frac{x}{1+t},\frac{y}{1+t},\frac{z}{1+t}\}\in B$ and is an isometry between $H^3$ and $B$. Let $P$ be a hyperplane in $H^3$ and $R$ be a reflection in $H^3$, then the corresponding hyperplane $\overline{P}$ in $B$ is 
\begin{equation}\label{conjugation1}
\overline{P}=\pi(P)
\end{equation}
and the corresponding reflection $\overline{R}$ in the Poincaré ball model $B$ is given by
\begin{equation}\label{conjugation2}
\overline{R}=\pi R\pi^{-1}
\end{equation}
All constructions in the hyperboloid model carry over to constructions in the Poincaré ball model via \eqref{conjugation1} and \eqref{conjugation2}.

\begin{figure}[h]
    \centering
    \includegraphics[width=0.4 \textwidth]{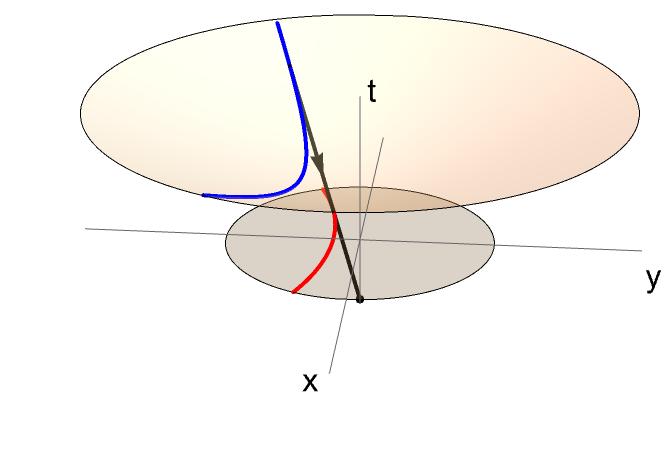}
    \caption{Projective relation between the 2D hyperboloid model and the 2D Poincaré disk.  The relation between the 3D hyperboloid model and the 3D Poinvaré ball follows similarly, although harder to visualize.}
    \label{projectiverelation}
\end{figure}

\section{Reference fundamental forms of flat surfaces and spheres in hyperbolic space}\label{APP:ab}
In this section we calculate reference fundamental forms of spheres and surfaces equidistant to a geodesic in hyperbolic space, and show that both of them can be reduced to the case of horospheres discussed in the main text.

\subsection{Spheres}
Let $r$ be the hyperbolic radius of a hyperbolic sphere centered at the origin of the Poincare ball, in the Poincaré ball model.  The sphere is described by $x^2+y^2+z^2=(\frac{e^r-1}{e^r+1})^2$. We use spherical coordinate $\{\theta,\phi\}$ to parametrize the sphere via
\begin{equation}
\begin{aligned}
&x=\frac{e^r-1}{e^r+1}\sin\theta\cos\phi\\
&y=\frac{e^r-1}{e^r+1}\sin\theta\sin\phi\\
&z=\frac{e^r-1}{e^r+1}\cos\theta\\
\end{aligned}
\end{equation}
The metric tensor under coordinate system $\{\theta,\phi\}$ is
\begin{equation}
ds^2=4\frac{dx^2+dy^2+dz^2}{(1-x^2-y^2-z^2)^2}=\sinh^2{r}(d\theta^2+\sin^2\theta d\phi^2)
\end{equation}
The reference fundamental forms are
\begin{equation}
\begin{aligned}
&\overline{a}=\begin{pmatrix}
\sinh^2{r} & 0 \\
0 & \sinh^2{r}\sin^2\theta \\
\end{pmatrix}\\
&\overline{b}=\begin{pmatrix}
-\cosh{r}\sinh{r} & 0 \\
0 & -\sin^2\theta\cosh{r}\sinh{r} \\
\end{pmatrix}\\
\end{aligned}
\end{equation}
In the large $r$ limit, near the equator $\theta=\frac{\pi}{2}$
\begin{equation}
\overline{a}\approx -\overline{b}\propto  \begin{pmatrix}
1 & 0 \\
0 & 1 \\
\end{pmatrix}
\end{equation}
reducing to the horosphere case.  Other parts of the sphere exhibit the same mode of geometric frustration given the rotational symmetry of the sphere, although they require a different choice of coordinate system to reduce to this particular form of the first and second reference fundamental forms.

\subsection{Surfaces equidistant to a geodesic}
A geodesic connecting points $p,q$ is the shortest curve joining $p,q$. We choose a geodesic $\gamma(t)=\{t,0,0,\sqrt{t^2-1}\}\in H^3$ in the hyperboloid model.  
Let $S$ be the surface consisting of points in $H^3$ of distance $r$ to the geodesic $\gamma$. 
We can parametrize $S$ using coordinate $\{\phi,h\}$ by
\begin{equation}
\begin{aligned}
&t=\cosh{r}\cosh{h}\\
&x=\sinh{r}\cos{\phi}\\
&y=\sinh{r}\sin{\phi}\\
&z=\cosh{r}\sinh{h}\\
\end{aligned}
\end{equation}
The metric tensor $\overline{g}$ under coordinate system $\{\phi,h\}$ is
\begin{equation}
\begin{aligned}
ds^2&=-dt^2+dx^2+dy^2+dz^2\\
&=\frac{1+\cosh{2r}}{2}dh^2+\frac{-1+\cosh{2r}}{2}d\phi^2
\end{aligned}
\end{equation}
The reference fundamental forms are
\begin{equation}
\begin{aligned}
&\overline{a}=\begin{pmatrix}
\frac{1+\cosh{2r}}{2} & 0 \\
0 & \frac{-1+\cosh{2r}}{2} \\
\end{pmatrix}\\
&\overline{b}=\begin{pmatrix}
-\frac{\sinh{2r}}{2} & 0 \\
0 & -\frac{\sinh{2r}}{2} \\
\end{pmatrix}
\end{aligned}
\end{equation}
In the large $r$ limit, 
\begin{equation}
\overline{a}\approx -\overline{b}\propto  \begin{pmatrix}
1 & 0 \\
0 & 1 \\
\end{pmatrix}
\end{equation}
reducing to the horosphere case.

\section{Energy minimization and morphology of $\overline{M}$ in Euclidean space}\label{APP:miniminazation}
In this section, we show the minimization procedure of $E_{elastic}+E_{repulsion}$. For a disk of radius $r$ on the $\delta=0$ horosphere, since stereographic projection gives an isometry between the horosphere and a plane, the $\{\alpha,\beta\}$ coordinates of the disk of radius $r$ in the $\delta=0$ horosphere form a disk of the same radius in Euclidean space. The geometric frustration, however, is reflected in the reference second fundamental form $\overline{b}$ where the next layer (small increment of $\delta$) shrinks relative to the $\delta=0$ horosphere.


We triangulate the disk for computational minimization of the energy. For a given embedding $F$ from the nodes of the triangulation into $E^3$, we use the method in reference \cite{https://doi.org/10.1111/j.1467-8659.2006.00974.x} to compute its fundamental forms and set $h=R$, $\nu=0$ to calculate the elastic energy. We also put a total charge of $Q=\frac{8\sqrt{\pi Y\epsilon_0}r^2}{9}$ (the coefficient is choosen so that the elastic energy is of the same order as the repulsion energy) on the entire disk, evenly distribute these charge to each node and use 
\begin{equation}
V(r_1,r_2)=\frac{1}{4\pi\epsilon_0}\frac{q_1q_2}{|r_1-r_2|}e^{-\kappa|r_1-r_2|}
\end{equation}
to calculate the repulsion energy under the embedding $\phi$. In this way, $E_{elastic}+E_{repulsion}$ is a function of nodes positions in $\mathbb{R}^3$. we then do gradient descent $E_{elastic}+E_{repulsion}$ with respect to nodes positions starting from many different initial configurations and pick the final configuration with minimal energy. After minimization, for radius $r=5R$ we obtained the stretching energy $E_{strechting}/E_{repulsion}=0.082$, the bending energy $E_{bending}/E_{repulsion}=0.661$ and the elastic energy $E_{elastic}/E_{repulsion}=0.743$.

After obtaining the morphology of the mid-surface, we reconstruct the discrete assembly of icosahedra on $\overline{M}$ using the following method. There are 31 icosahedra intersecting with the disk for the case of radius 5 and we project the centers of the 31 icosahedra onto the $\delta=0$ horosphere along the $\delta$ line, obtaining their corresponding $\{\alpha,\beta\}$ coordinates and find these points (\{$p_1,...p_{31}$\}) on the embedding surface in $E^3$. Then we calculate the hyperbolic distance between the centers of the 31 icosahedra to the $\delta=0$ horosphere respectively (\{$d_1,...,d_{31}$\}). Next, we place Euclidean icosahedra of the same size as the hyperbolic icosahedra (in terms of the body-center to face-center distance)
to the Euclidean space such that  
the $i^{th}$ icosahedron lies above/below the point $p_{i}$ on midsurface by distance $d_i$.
Finally, we rotate the icosahedra so that the overlapping among them is the smallest.  The result is shown in Fig.~\ref{morphology}h.

\bibliography{GFA}

\providecommand{\noopsort}[1]{}\providecommand{\singleletter}[1]{#1}%
\begin{thebibliography}{29}%
\makeatletter
\providecommand \@ifxundefined [1]{%
 \@ifx{#1\undefined}
}%
\providecommand \@ifnum [1]{%
 \ifnum #1\expandafter \@firstoftwo
 \else \expandafter \@secondoftwo
 \fi
}%
\providecommand \@ifx [1]{%
 \ifx #1\expandafter \@firstoftwo
 \else \expandafter \@secondoftwo
 \fi
}%
\providecommand \natexlab [1]{#1}%
\providecommand \enquote  [1]{``#1''}%
\providecommand \bibnamefont  [1]{#1}%
\providecommand \bibfnamefont [1]{#1}%
\providecommand \citenamefont [1]{#1}%
\providecommand \href@noop [0]{\@secondoftwo}%
\providecommand \href [0]{\begingroup \@sanitize@url \@href}%
\providecommand \@href[1]{\@@startlink{#1}\@@href}%
\providecommand \@@href[1]{\endgroup#1\@@endlink}%
\providecommand \@sanitize@url [0]{\catcode `\\12\catcode `\$12\catcode
  `\&12\catcode `\#12\catcode `\^12\catcode `\_12\catcode `\%12\relax}%
\providecommand \@@startlink[1]{}%
\providecommand \@@endlink[0]{}%
\providecommand \url  [0]{\begingroup\@sanitize@url \@url }%
\providecommand \@url [1]{\endgroup\@href {#1}{\urlprefix }}%
\providecommand \urlprefix  [0]{URL }%
\providecommand \Eprint [0]{\href }%
\providecommand \doibase [0]{http://dx.doi.org/}%
\providecommand \selectlanguage [0]{\@gobble}%
\providecommand \bibinfo  [0]{\@secondoftwo}%
\providecommand \bibfield  [0]{\@secondoftwo}%
\providecommand \translation [1]{[#1]}%
\providecommand \BibitemOpen [0]{}%
\providecommand \bibitemStop [0]{}%
\providecommand \bibitemNoStop [0]{.\EOS\space}%
\providecommand \EOS [0]{\spacefactor3000\relax}%
\providecommand \BibitemShut  [1]{\csname bibitem#1\endcsname}%
\let\auto@bib@innerbib\@empty
\bibitem [{\citenamefont {Sadoc}\ and\ \citenamefont
  {Mosseri}(1999)}]{sadoc_mosseri_1999}%
  \BibitemOpen
  \bibfield  {author} {\bibinfo {author} {\bibfnamefont {J.-F.}\ \bibnamefont
  {Sadoc}}\ and\ \bibinfo {author} {\bibfnamefont {R.}~\bibnamefont
  {Mosseri}},\ }\href@noop {} {\emph {\bibinfo {title} {Geometrical
  Frustration}}},\ Collection Alea-Saclay: Monographs and Texts in Statistical
  Physics\ (\bibinfo  {publisher} {Cambridge University Press},\ \bibinfo
  {year} {1999})\BibitemShut {NoStop}%
\bibitem [{\citenamefont {Bruss}\ and\ \citenamefont
  {Grason}(2012)}]{bruss2012non}%
  \BibitemOpen
  \bibfield  {author} {\bibinfo {author} {\bibfnamefont {I.~R.}\ \bibnamefont
  {Bruss}}\ and\ \bibinfo {author} {\bibfnamefont {G.~M.}\ \bibnamefont
  {Grason}},\ }\href@noop {} {\bibfield  {journal} {\bibinfo  {journal}
  {Proceedings of the National Academy of Sciences}\ }\textbf {\bibinfo
  {volume} {109}},\ \bibinfo {pages} {10781} (\bibinfo {year}
  {2012})}\BibitemShut {NoStop}%
\bibitem [{\citenamefont {Irvine}\ \emph {et~al.}(2010)\citenamefont {Irvine},
  \citenamefont {Vitelli},\ and\ \citenamefont {Chaikin}}]{irvine2010pleats}%
  \BibitemOpen
  \bibfield  {author} {\bibinfo {author} {\bibfnamefont {W.~T.}\ \bibnamefont
  {Irvine}}, \bibinfo {author} {\bibfnamefont {V.}~\bibnamefont {Vitelli}}, \
  and\ \bibinfo {author} {\bibfnamefont {P.~M.}\ \bibnamefont {Chaikin}},\
  }\href@noop {} {\bibfield  {journal} {\bibinfo  {journal} {Nature}\ }\textbf
  {\bibinfo {volume} {468}},\ \bibinfo {pages} {947} (\bibinfo {year}
  {2010})}\BibitemShut {NoStop}%
\bibitem [{\citenamefont {Grason}(2016)}]{grason2016perspective}%
  \BibitemOpen
  \bibfield  {author} {\bibinfo {author} {\bibfnamefont {G.~M.}\ \bibnamefont
  {Grason}},\ }\href@noop {} {\bibfield  {journal} {\bibinfo  {journal} {The
  Journal of Chemical Physics}\ }\textbf {\bibinfo {volume} {145}},\ \bibinfo
  {pages} {110901} (\bibinfo {year} {2016})}\BibitemShut {NoStop}%
\bibitem [{\citenamefont {Lenz}\ and\ \citenamefont
  {Witten}(2017)}]{lenz2017geometrical}%
  \BibitemOpen
  \bibfield  {author} {\bibinfo {author} {\bibfnamefont {M.}~\bibnamefont
  {Lenz}}\ and\ \bibinfo {author} {\bibfnamefont {T.~A.}\ \bibnamefont
  {Witten}},\ }\href@noop {} {\bibfield  {journal} {\bibinfo  {journal} {Nature
  physics}\ }\textbf {\bibinfo {volume} {13}},\ \bibinfo {pages} {1100}
  (\bibinfo {year} {2017})}\BibitemShut {NoStop}%
\bibitem [{\citenamefont {Travesset}(2017)}]{travesset2017nanoparticle}%
  \BibitemOpen
  \bibfield  {author} {\bibinfo {author} {\bibfnamefont {A.}~\bibnamefont
  {Travesset}},\ }\href@noop {} {\bibfield  {journal} {\bibinfo  {journal}
  {Physical Review Letters}\ }\textbf {\bibinfo {volume} {119}},\ \bibinfo
  {pages} {115701} (\bibinfo {year} {2017})}\BibitemShut {NoStop}%
\bibitem [{\citenamefont {Haddad}\ \emph {et~al.}(2019)\citenamefont {Haddad},
  \citenamefont {Aharoni}, \citenamefont {Sharon}, \citenamefont {Shtukenberg},
  \citenamefont {Kahr},\ and\ \citenamefont {Efrati}}]{haddad2019twist}%
  \BibitemOpen
  \bibfield  {author} {\bibinfo {author} {\bibfnamefont {A.}~\bibnamefont
  {Haddad}}, \bibinfo {author} {\bibfnamefont {H.}~\bibnamefont {Aharoni}},
  \bibinfo {author} {\bibfnamefont {E.}~\bibnamefont {Sharon}}, \bibinfo
  {author} {\bibfnamefont {A.~G.}\ \bibnamefont {Shtukenberg}}, \bibinfo
  {author} {\bibfnamefont {B.}~\bibnamefont {Kahr}}, \ and\ \bibinfo {author}
  {\bibfnamefont {E.}~\bibnamefont {Efrati}},\ }\href@noop {} {\bibfield
  {journal} {\bibinfo  {journal} {Soft Matter}\ }\textbf {\bibinfo {volume}
  {15}},\ \bibinfo {pages} {116} (\bibinfo {year} {2019})}\BibitemShut
  {NoStop}%
\bibitem [{\citenamefont {Sadoc}\ \emph {et~al.}(2020)\citenamefont {Sadoc},
  \citenamefont {Mosseri},\ and\ \citenamefont {Selinger}}]{sadoc2020liquid}%
  \BibitemOpen
  \bibfield  {author} {\bibinfo {author} {\bibfnamefont {J.-F.}\ \bibnamefont
  {Sadoc}}, \bibinfo {author} {\bibfnamefont {R.}~\bibnamefont {Mosseri}}, \
  and\ \bibinfo {author} {\bibfnamefont {J.~V.}\ \bibnamefont {Selinger}},\
  }\href@noop {} {\bibfield  {journal} {\bibinfo  {journal} {New Journal of
  Physics}\ }\textbf {\bibinfo {volume} {22}},\ \bibinfo {pages} {093036}
  (\bibinfo {year} {2020})}\BibitemShut {NoStop}%
\bibitem [{\citenamefont {Li}\ \emph {et~al.}(2020)\citenamefont {Li},
  \citenamefont {Shtukenberg}, \citenamefont {Vogt-Maranto}, \citenamefont
  {Efrati}, \citenamefont {Raiteri}, \citenamefont {Gale}, \citenamefont
  {Rohl},\ and\ \citenamefont {Kahr}}]{li2020some}%
  \BibitemOpen
  \bibfield  {author} {\bibinfo {author} {\bibfnamefont {C.}~\bibnamefont
  {Li}}, \bibinfo {author} {\bibfnamefont {A.~G.}\ \bibnamefont {Shtukenberg}},
  \bibinfo {author} {\bibfnamefont {L.}~\bibnamefont {Vogt-Maranto}}, \bibinfo
  {author} {\bibfnamefont {E.}~\bibnamefont {Efrati}}, \bibinfo {author}
  {\bibfnamefont {P.}~\bibnamefont {Raiteri}}, \bibinfo {author} {\bibfnamefont
  {J.~D.}\ \bibnamefont {Gale}}, \bibinfo {author} {\bibfnamefont {A.~L.}\
  \bibnamefont {Rohl}}, \ and\ \bibinfo {author} {\bibfnamefont
  {B.}~\bibnamefont {Kahr}},\ }\href@noop {} {\bibfield  {journal} {\bibinfo
  {journal} {The Journal of Physical Chemistry C}\ }\textbf {\bibinfo {volume}
  {124}},\ \bibinfo {pages} {15616} (\bibinfo {year} {2020})}\BibitemShut
  {NoStop}%
\bibitem [{\citenamefont {Meiri}\ and\ \citenamefont
  {Efrati}(2021)}]{meiri2021cumulative}%
  \BibitemOpen
  \bibfield  {author} {\bibinfo {author} {\bibfnamefont {S.}~\bibnamefont
  {Meiri}}\ and\ \bibinfo {author} {\bibfnamefont {E.}~\bibnamefont {Efrati}},\
  }\href@noop {} {\bibfield  {journal} {\bibinfo  {journal} {Physical Review
  E}\ }\textbf {\bibinfo {volume} {104}},\ \bibinfo {pages} {054601} (\bibinfo
  {year} {2021})}\BibitemShut {NoStop}%
\bibitem [{\citenamefont {Serafin}\ \emph {et~al.}(2021)\citenamefont
  {Serafin}, \citenamefont {Lu}, \citenamefont {Kotov}, \citenamefont {Sun},\
  and\ \citenamefont {Mao}}]{serafin2021frustrated}%
  \BibitemOpen
  \bibfield  {author} {\bibinfo {author} {\bibfnamefont {F.}~\bibnamefont
  {Serafin}}, \bibinfo {author} {\bibfnamefont {J.}~\bibnamefont {Lu}},
  \bibinfo {author} {\bibfnamefont {N.}~\bibnamefont {Kotov}}, \bibinfo
  {author} {\bibfnamefont {K.}~\bibnamefont {Sun}}, \ and\ \bibinfo {author}
  {\bibfnamefont {X.}~\bibnamefont {Mao}},\ }\href@noop {} {\bibfield
  {journal} {\bibinfo  {journal} {Nature Communications}\ }\textbf {\bibinfo
  {volume} {12}},\ \bibinfo {pages} {4925} (\bibinfo {year}
  {2021})}\BibitemShut {NoStop}%
\bibitem [{\citenamefont {Hall}\ \emph {et~al.}(2023)\citenamefont {Hall},
  \citenamefont {Stevens},\ and\ \citenamefont {Grason}}]{hall2023building}%
  \BibitemOpen
  \bibfield  {author} {\bibinfo {author} {\bibfnamefont {D.~M.}\ \bibnamefont
  {Hall}}, \bibinfo {author} {\bibfnamefont {M.~J.}\ \bibnamefont {Stevens}}, \
  and\ \bibinfo {author} {\bibfnamefont {G.~M.}\ \bibnamefont {Grason}},\
  }\href@noop {} {\bibfield  {journal} {\bibinfo  {journal} {Soft Matter}\
  }\textbf {\bibinfo {volume} {19}},\ \bibinfo {pages} {858} (\bibinfo {year}
  {2023})}\BibitemShut {NoStop}%
\bibitem [{\citenamefont {Hackney}\ \emph {et~al.}(2023)\citenamefont
  {Hackney}, \citenamefont {Amey},\ and\ \citenamefont
  {Grason}}]{hackney2023dispersed}%
  \BibitemOpen
  \bibfield  {author} {\bibinfo {author} {\bibfnamefont {N.~W.}\ \bibnamefont
  {Hackney}}, \bibinfo {author} {\bibfnamefont {C.}~\bibnamefont {Amey}}, \
  and\ \bibinfo {author} {\bibfnamefont {G.~M.}\ \bibnamefont {Grason}},\
  }\href@noop {} {\bibfield  {journal} {\bibinfo  {journal} {arXiv preprint
  arXiv:2303.02121}\ } (\bibinfo {year} {2023})}\BibitemShut {NoStop}%
\bibitem [{\citenamefont {Modes}\ and\ \citenamefont
  {Kamien}(2007)}]{modes2007hard}%
  \BibitemOpen
  \bibfield  {author} {\bibinfo {author} {\bibfnamefont {C.~D.}\ \bibnamefont
  {Modes}}\ and\ \bibinfo {author} {\bibfnamefont {R.~D.}\ \bibnamefont
  {Kamien}},\ }\href@noop {} {\bibfield  {journal} {\bibinfo  {journal}
  {Physical review letters}\ }\textbf {\bibinfo {volume} {99}},\ \bibinfo
  {pages} {235701} (\bibinfo {year} {2007})}\BibitemShut {NoStop}%
\bibitem [{\citenamefont {Ackerman}\ and\ \citenamefont
  {Smalyukh}(2017)}]{ackerman2017diversity}%
  \BibitemOpen
  \bibfield  {author} {\bibinfo {author} {\bibfnamefont {P.~J.}\ \bibnamefont
  {Ackerman}}\ and\ \bibinfo {author} {\bibfnamefont {I.~I.}\ \bibnamefont
  {Smalyukh}},\ }\href@noop {} {\bibfield  {journal} {\bibinfo  {journal}
  {Physical Review X}\ }\textbf {\bibinfo {volume} {7}},\ \bibinfo {pages}
  {011006} (\bibinfo {year} {2017})}\BibitemShut {NoStop}%
\bibitem [{\citenamefont {Sch{\"o}nh{\"o}fer}\ \emph
  {et~al.}(2023)\citenamefont {Sch{\"o}nh{\"o}fer}, \citenamefont {Sun},
  \citenamefont {Mao},\ and\ \citenamefont
  {Glotzer}}]{schonhofer2023rationalizing}%
  \BibitemOpen
  \bibfield  {author} {\bibinfo {author} {\bibfnamefont {P.~W.}\ \bibnamefont
  {Sch{\"o}nh{\"o}fer}}, \bibinfo {author} {\bibfnamefont {K.}~\bibnamefont
  {Sun}}, \bibinfo {author} {\bibfnamefont {X.}~\bibnamefont {Mao}}, \ and\
  \bibinfo {author} {\bibfnamefont {S.~C.}\ \bibnamefont {Glotzer}},\
  }\href@noop {} {\bibfield  {journal} {\bibinfo  {journal} {arXiv preprint
  arXiv:2305.07786}\ } (\bibinfo {year} {2023})}\BibitemShut {NoStop}%
\bibitem [{\citenamefont {Hofmeister}(1998)}]{hofmeister1998forty}%
  \BibitemOpen
  \bibfield  {author} {\bibinfo {author} {\bibfnamefont {H.}~\bibnamefont
  {Hofmeister}},\ }\href@noop {} {\bibfield  {journal} {\bibinfo  {journal}
  {Crystal Research and Technology: Journal of Experimental and Industrial
  Crystallography}\ }\textbf {\bibinfo {volume} {33}},\ \bibinfo {pages} {3}
  (\bibinfo {year} {1998})}\BibitemShut {NoStop}%
\bibitem [{\citenamefont {De~Nijs}\ \emph {et~al.}(2015)\citenamefont
  {De~Nijs}, \citenamefont {Dussi}, \citenamefont {Smallenburg}, \citenamefont
  {Meeldijk}, \citenamefont {Groenendijk}, \citenamefont {Filion},
  \citenamefont {Imhof}, \citenamefont {Van~Blaaderen},\ and\ \citenamefont
  {Dijkstra}}]{de2015entropy}%
  \BibitemOpen
  \bibfield  {author} {\bibinfo {author} {\bibfnamefont {B.}~\bibnamefont
  {De~Nijs}}, \bibinfo {author} {\bibfnamefont {S.}~\bibnamefont {Dussi}},
  \bibinfo {author} {\bibfnamefont {F.}~\bibnamefont {Smallenburg}}, \bibinfo
  {author} {\bibfnamefont {J.~D.}\ \bibnamefont {Meeldijk}}, \bibinfo {author}
  {\bibfnamefont {D.~J.}\ \bibnamefont {Groenendijk}}, \bibinfo {author}
  {\bibfnamefont {L.}~\bibnamefont {Filion}}, \bibinfo {author} {\bibfnamefont
  {A.}~\bibnamefont {Imhof}}, \bibinfo {author} {\bibfnamefont
  {A.}~\bibnamefont {Van~Blaaderen}}, \ and\ \bibinfo {author} {\bibfnamefont
  {M.}~\bibnamefont {Dijkstra}},\ }\href@noop {} {\bibfield  {journal}
  {\bibinfo  {journal} {Nature materials}\ }\textbf {\bibinfo {volume} {14}},\
  \bibinfo {pages} {56} (\bibinfo {year} {2015})}\BibitemShut {NoStop}%
\bibitem [{\citenamefont {Yan}\ \emph {et~al.}(2019)\citenamefont {Yan},
  \citenamefont {Feng}, \citenamefont {Kim}, \citenamefont {Lu}, \citenamefont
  {Kumar}, \citenamefont {Mu}, \citenamefont {Wu}, \citenamefont {Mao},\ and\
  \citenamefont {Kotov}}]{yan2019self}%
  \BibitemOpen
  \bibfield  {author} {\bibinfo {author} {\bibfnamefont {J.}~\bibnamefont
  {Yan}}, \bibinfo {author} {\bibfnamefont {W.}~\bibnamefont {Feng}}, \bibinfo
  {author} {\bibfnamefont {J.-Y.}\ \bibnamefont {Kim}}, \bibinfo {author}
  {\bibfnamefont {J.}~\bibnamefont {Lu}}, \bibinfo {author} {\bibfnamefont
  {P.}~\bibnamefont {Kumar}}, \bibinfo {author} {\bibfnamefont
  {Z.}~\bibnamefont {Mu}}, \bibinfo {author} {\bibfnamefont {X.}~\bibnamefont
  {Wu}}, \bibinfo {author} {\bibfnamefont {X.}~\bibnamefont {Mao}}, \ and\
  \bibinfo {author} {\bibfnamefont {N.~A.}\ \bibnamefont {Kotov}},\ }\href@noop
  {} {\bibfield  {journal} {\bibinfo  {journal} {Chemistry of Materials}\
  }\textbf {\bibinfo {volume} {32}},\ \bibinfo {pages} {476} (\bibinfo {year}
  {2019})}\BibitemShut {NoStop}%
\bibitem [{\citenamefont {Koll{\'a}r}\ \emph {et~al.}(2019)\citenamefont
  {Koll{\'a}r}, \citenamefont {Fitzpatrick},\ and\ \citenamefont
  {Houck}}]{kollar2019hyperbolic}%
  \BibitemOpen
  \bibfield  {author} {\bibinfo {author} {\bibfnamefont {A.~J.}\ \bibnamefont
  {Koll{\'a}r}}, \bibinfo {author} {\bibfnamefont {M.}~\bibnamefont
  {Fitzpatrick}}, \ and\ \bibinfo {author} {\bibfnamefont {A.~A.}\ \bibnamefont
  {Houck}},\ }\href@noop {} {\bibfield  {journal} {\bibinfo  {journal}
  {Nature}\ }\textbf {\bibinfo {volume} {571}},\ \bibinfo {pages} {45}
  (\bibinfo {year} {2019})}\BibitemShut {NoStop}%
\bibitem [{\citenamefont {Maciejko}\ and\ \citenamefont
  {Rayan}(2021)}]{maciejko2021hyperbolic}%
  \BibitemOpen
  \bibfield  {author} {\bibinfo {author} {\bibfnamefont {J.}~\bibnamefont
  {Maciejko}}\ and\ \bibinfo {author} {\bibfnamefont {S.}~\bibnamefont
  {Rayan}},\ }\href@noop {} {\bibfield  {journal} {\bibinfo  {journal} {Science
  advances}\ }\textbf {\bibinfo {volume} {7}},\ \bibinfo {pages} {eabe9170}
  (\bibinfo {year} {2021})}\BibitemShut {NoStop}%
\bibitem [{\citenamefont {Yu}\ \emph {et~al.}(2020)\citenamefont {Yu},
  \citenamefont {Piao},\ and\ \citenamefont {Park}}]{yu2020topological}%
  \BibitemOpen
  \bibfield  {author} {\bibinfo {author} {\bibfnamefont {S.}~\bibnamefont
  {Yu}}, \bibinfo {author} {\bibfnamefont {X.}~\bibnamefont {Piao}}, \ and\
  \bibinfo {author} {\bibfnamefont {N.}~\bibnamefont {Park}},\ }\href@noop {}
  {\bibfield  {journal} {\bibinfo  {journal} {Physical Review Letters}\
  }\textbf {\bibinfo {volume} {125}},\ \bibinfo {pages} {053901} (\bibinfo
  {year} {2020})}\BibitemShut {NoStop}%
\bibitem [{\citenamefont {Cheng}\ \emph {et~al.}(2022)\citenamefont {Cheng},
  \citenamefont {Serafin}, \citenamefont {McInerney}, \citenamefont {Rocklin},
  \citenamefont {Sun},\ and\ \citenamefont {Mao}}]{cheng2022band}%
  \BibitemOpen
  \bibfield  {author} {\bibinfo {author} {\bibfnamefont {N.}~\bibnamefont
  {Cheng}}, \bibinfo {author} {\bibfnamefont {F.}~\bibnamefont {Serafin}},
  \bibinfo {author} {\bibfnamefont {J.}~\bibnamefont {McInerney}}, \bibinfo
  {author} {\bibfnamefont {Z.}~\bibnamefont {Rocklin}}, \bibinfo {author}
  {\bibfnamefont {K.}~\bibnamefont {Sun}}, \ and\ \bibinfo {author}
  {\bibfnamefont {X.}~\bibnamefont {Mao}},\ }\href@noop {} {\bibfield
  {journal} {\bibinfo  {journal} {Physical Review Letters}\ }\textbf {\bibinfo
  {volume} {129}},\ \bibinfo {pages} {088002} (\bibinfo {year}
  {2022})}\BibitemShut {NoStop}%
\bibitem [{\citenamefont {Maciejko}\ and\ \citenamefont
  {Rayan}(2022)}]{maciejko2022automorphic}%
  \BibitemOpen
  \bibfield  {author} {\bibinfo {author} {\bibfnamefont {J.}~\bibnamefont
  {Maciejko}}\ and\ \bibinfo {author} {\bibfnamefont {S.}~\bibnamefont
  {Rayan}},\ }\href@noop {} {\bibfield  {journal} {\bibinfo  {journal}
  {Proceedings of the National Academy of Sciences}\ }\textbf {\bibinfo
  {volume} {119}},\ \bibinfo {pages} {e2116869119} (\bibinfo {year}
  {2022})}\BibitemShut {NoStop}%
\bibitem [{\citenamefont {Urwyler}\ \emph {et~al.}(2022)\citenamefont
  {Urwyler}, \citenamefont {Lenggenhager}, \citenamefont {Boettcher},
  \citenamefont {Thomale}, \citenamefont {Neupert},\ and\ \citenamefont
  {Bzdu{\v{s}}ek}}]{urwyler2022hyperbolic}%
  \BibitemOpen
  \bibfield  {author} {\bibinfo {author} {\bibfnamefont {D.~M.}\ \bibnamefont
  {Urwyler}}, \bibinfo {author} {\bibfnamefont {P.~M.}\ \bibnamefont
  {Lenggenhager}}, \bibinfo {author} {\bibfnamefont {I.}~\bibnamefont
  {Boettcher}}, \bibinfo {author} {\bibfnamefont {R.}~\bibnamefont {Thomale}},
  \bibinfo {author} {\bibfnamefont {T.}~\bibnamefont {Neupert}}, \ and\
  \bibinfo {author} {\bibfnamefont {T.}~\bibnamefont {Bzdu{\v{s}}ek}},\
  }\href@noop {} {\bibfield  {journal} {\bibinfo  {journal} {Physical Review
  Letters}\ }\textbf {\bibinfo {volume} {129}},\ \bibinfo {pages} {246402}
  (\bibinfo {year} {2022})}\BibitemShut {NoStop}%
\bibitem [{\citenamefont {Bzdu{\v{s}}ek}\ and\ \citenamefont
  {Maciejko}(2022)}]{bzduvsek2022flat}%
  \BibitemOpen
  \bibfield  {author} {\bibinfo {author} {\bibfnamefont {T.}~\bibnamefont
  {Bzdu{\v{s}}ek}}\ and\ \bibinfo {author} {\bibfnamefont {J.}~\bibnamefont
  {Maciejko}},\ }\href@noop {} {\bibfield  {journal} {\bibinfo  {journal}
  {Physical Review B}\ }\textbf {\bibinfo {volume} {106}},\ \bibinfo {pages}
  {155146} (\bibinfo {year} {2022})}\BibitemShut {NoStop}%
\bibitem [{\citenamefont {Coxeter}\ and\ \citenamefont
  {Coxeter}(1999)}]{coxeter1999beauty}%
  \BibitemOpen
  \bibfield  {author} {\bibinfo {author} {\bibfnamefont {H.}~\bibnamefont
  {Coxeter}}\ and\ \bibinfo {author} {\bibfnamefont {H.}~\bibnamefont
  {Coxeter}},\ }\href {https://books.google.com/books?id=p4o-Uf-i-IUC} {\emph
  {\bibinfo {title} {The Beauty of Geometry: Twelve Essays}}},\ Dover books on
  mathematics\ (\bibinfo  {publisher} {Dover Publications},\ \bibinfo {year}
  {1999})\BibitemShut {NoStop}%
\bibitem [{\citenamefont {Iversen}(1992)}]{iversen1992hyperbolic}%
  \BibitemOpen
  \bibfield  {author} {\bibinfo {author} {\bibfnamefont {B.}~\bibnamefont
  {Iversen}},\ }\href {https://books.google.com/books?id=Zlo\_StuX2t0C} {\emph
  {\bibinfo {title} {Hyperbolic Geometry}}},\ London Mathematical Society
  Student Texts\ (\bibinfo  {publisher} {Cambridge University Press},\ \bibinfo
  {year} {1992})\BibitemShut {NoStop}%
\bibitem [{\citenamefont {Grinspun}\ \emph {et~al.}(2006)\citenamefont
  {Grinspun}, \citenamefont {Gingold}, \citenamefont {Reisman},\ and\
  \citenamefont {Zorin}}]{https://doi.org/10.1111/j.1467-8659.2006.00974.x}%
  \BibitemOpen
  \bibfield  {author} {\bibinfo {author} {\bibfnamefont {E.}~\bibnamefont
  {Grinspun}}, \bibinfo {author} {\bibfnamefont {Y.}~\bibnamefont {Gingold}},
  \bibinfo {author} {\bibfnamefont {J.}~\bibnamefont {Reisman}}, \ and\
  \bibinfo {author} {\bibfnamefont {D.}~\bibnamefont {Zorin}},\ }\href
  {\doibase https://doi.org/10.1111/j.1467-8659.2006.00974.x} {\bibfield
  {journal} {\bibinfo  {journal} {Computer Graphics Forum}\ }\textbf {\bibinfo
  {volume} {25}},\ \bibinfo {pages} {547} (\bibinfo {year} {2006})},\ \Eprint
  {http://arxiv.org/abs/https://onlinelibrary.wiley.com/doi/pdf/10.1111/j.1467-8659.2006.00974.x}
  {https://onlinelibrary.wiley.com/doi/pdf/10.1111/j.1467-8659.2006.00974.x}
  \BibitemShut {NoStop}%
\end{thebibliography}%

\end{document}